\documentclass[aps,prb,twocolumn]{revtex4}
\pdfoutput = 1

\usepackage{graphicx}
\usepackage{amsfonts}
\usepackage{amssymb}
\usepackage{amsmath}

\usepackage{color}

\font\elevenmib=cmmib10 scaled 1095

 \skewchar\elevenmib='177
\newfam\mibfam

\renewcommand{\Re}{{\mathrm{Re}}\, }
\renewcommand{\Im}{{\mathrm{Im}}\, }
\newcommand{\beq}{
\begin{equation}
	}
	\newcommand{\eeq}{
\end{equation}
}

\newcommand{\barr}{
\begin{eqnarray}
	}
	\newcommand{\earr}{
\end{eqnarray}
}

 \mathchardef\sigma="711B

\def\abs#1{\mathopen| #1 \mathclose|}

\begin{document}

\title{There are no Goldstone bosons on the Bethe lattice}

\author{C. R. Laumann}

\affiliation{Department of Physics, Joseph Henry Laboratories, Princeton University, Princeton NJ 08544}

\author{S. A. Parameswaran}

\affiliation{Department of Physics, Joseph Henry Laboratories, Princeton University, Princeton NJ 08544}

\author{S. L. Sondhi}

\affiliation{Department of Physics, Joseph Henry Laboratories, Princeton University, Princeton NJ 08544}

\date{\today}

\begin{abstract}
	We discuss symmetry breaking quantum phase transitions on the oft studied Bethe lattice in the context of the ferromagnetic scalar spherical model or, equivalently, the infinite $N_f$ limit of ferromagnetic models with $O(N_f)$ symmetry.
	We show that the approach to quantum criticality is characterized by the vanishing of a gap to just the global modes so that {\it all} local correlation functions continue to exhibit massive behavior. This behavior persists into the broken symmetry phase even as the order parameter develops an expectation value and thus there are no massless Goldstone bosons in the spectrum.
	We relate this feature to a spectral property of the graph Laplacian shared by the set of ``expander'' graphs, and argue that our results apply to symmetry breaking transitions on such graphs quite generally.
\end{abstract}

\maketitle

\section{Introduction}
\label{sec:Intro}

The study of statistical mechanical systems on the Bethe lattice dates back to Bethe's early work on binary alloys in the 1930s\cite{Bethe:1935qc}.
Formally, the Bethe lattice is a regular graph with no loops in the infinite size limit; its tree-like local structure makes it amenable to real space self-consistent treatments for systems with short ranged interactions without entirely losing the notion of distance, as happens if instead we resort to infinite ranged interactions. Specifically, for ferromagnetic, and hence unfrustrated, models it makes the Bethe-Peierls approximation exact. For frustrated systems, such as spin glasses, the simplification achieved is more problematic and has been the subject of much ongoing work\cite{Mezard:2001p84}. Altogether, there is a large set of results on various aspects of classical statistical mechanics on the Bethe lattice.

Quantum statistical mechanics on the Bethe lattice has received less attention by comparison. One early set of results concern the Hubbard model on the Bethe lattice in the infinite coordination limit \cite{Georges:1996hl}. More recently, purely bosonic models on the Bethe lattice have been the subject of investigation with the introduction of the quantum cavity method. \cite{Laumann:2008rw,Semerjian:2009uo,Krzakala:2008rc} This has led to computational results on the quantum Ising model, the quantum Ising spin glass and on the Bose-Hubbard model. In addition, we note variational work on the ground state of the quantum Ising model \cite{Nagaj:2008xe} and on a spherical model of a spin glass \cite{Kopec:2006qr}.

In this paper we study the quantum unfrustrated (ferromagnetic) problem on the Bethe lattice analytically. For technical reasons, we do this nominally in the form of the spherical model for a scalar field but it is essentially also the large $N_f$ limit of the nearest neighbor $O(N_f)$ quantum rotor model on the Bethe lattice. We find some surprising results from the perspective of quantum phase transitions on Euclidean lattices. First, we find that the critical point is marked by a {\it single} global mode (or $N_f$ modes in the $O(N_f)$ interpretation) descending in energy to become degenerate with the ground state while all other states remain a {\it finite} distance away in energy. Second, no local operators couple to this mode in the thermodynamic limit so that all local correlation functions exhibit a gap at criticality. Alternatively, once the thermodynamic limit has been taken, the response to a field on any finite subvolume is bounded even if that volume is taken to infinity.  Third, this state of affairs persists into the broken symmetry phase in that while there are 2 (or $O(N_f)$ worth of) degenerate broken symmetry states as required by the global symmetry, there are no Goldstone bosons and all local correlators remain gapped. We believe our results are robust to moving away from the spherical/$N_f=\infty$ limit. Indeed, the $N_f=2$ problem is the particle-hole symmetric transition in the Bose-Hubbard model studied using the cavity method by Semerjian \emph{et al}\cite{Semerjian:2009uo} whose results imply that a macroscopic superfluid density develops at the quantum phase transition (QPT) without closing the excitation gap in the single site correlation function.

These results should surprise the reader. On the face of it, the lack of Goldstone behavior confounds the standard intuition regarding broken symmetry phases. Aficionados of classical models on the Bethe lattice may counter, rightly, that their phase transitions are accompanied by exponential rather than power law correlations, as this is sufficient for the classical susceptibility to diverge. However, if one computes the quantum susceptibility -- by which we mean the response to an infinitesimal field applied over a subvolume whose size is taken to infinity after the thermodynamic limit -- it remains finite, because the matrix elements in question do not couple to the global mode. The analogous susceptibility of the classical model diverges to signal the phase transition.

We turn now to a brief discussion of what we mean by the Bethe lattice as it plays an important role in the following. The traditional, and simplest, way to define the Bethe lattice is through a sequence of Cayley trees. The Bethe lattice then corresponds to the interior of a large tree where each site is connected to exactly $z$ neighbors. This construction makes clear that the number of sites within a fixed distance of a given site scales exponentially with distance and thus connects, qualitatively, with the infinite dimensional limit of hypercubic lattices that is also often invoked in the context of making self-consistent treatments exact. Unfortunately, the boundary of a finite Cayley tree is always a finite fraction of the bulk whence the choice of boundary conditions can complicate the thermodynamic limit\cite{Eggarter:1974kx}. Indeed, the difference between a ferromagnet and a glass can be phrased \emph{solely} in terms of boundary conditions.
To get around this problem, one can proceed differently and consider instead the ensemble of $z-$regular graphs with $N$ nodes. This consists of all graphs where each site is connected to exactly $z$ nearest neighbors. Picking a graph at random from this ensemble in the $N \rightarrow \infty$ limit yields an alternative definition of the Bethe lattice which is known to be perfectly satisfactory for the classical ferromagnetic problem\cite{Johnston:1998fv} and that is what we shall use in this paper. At finite $N$, members of the ensemble contain loops of characteristic size $\log N$ and the graphs are not entirely homogeneous. However, homogeneity is restored in the limit $N \rightarrow \infty$ as the loop size diverges.

With this definition in hand, we can comment on the feature of the Bethe lattice that brings about the surprising features we report: the boundary of any subvolume of the lattice is always a finite fraction of the whole whence (arbitrarily) localized excitations cannot be made low energy. In more formal terms, the Bethe lattice has what is known to graph theorists as a positive Cheeger constant $h$ -- that is the minimal ratio of boundary to bulk of its nontrivial subgraphs. A theorem -- the Cheeger bound -- guarantees that the Laplacian has a gap of at least $h^2/2$ between the uniform Perron-Frobenius ground state and the rest of the spectrum. Moreover, this property is shared by a large collection of so-called expander graphs\cite{Hoory:2006qd}, of much interest in quantum and classical information theory, and we expect our qualitative results to be shared by models on any such graph. 

In the balance of this paper, we first introduce and solve the classical spherical model on the Bethe lattice which introduces most of the ingredients needed in a simpler setting. This classical model on the Bethe lattice has been solved previously\cite{Cassi:1990qq} by a different technique. We next introduce and solve the quantum spherical model which exhibits the features we described above. (We note that the quantum spherical model on Euclidean lattices has been solved by Vojta\cite{Vojta:1996gd}.) We end with a discussion and comments on the generality of our results.

\section{The Classical Spherical Model}
\label{sec:Classical_Model}

The classical spherical model on a $z$-regular graph $\mathcal{G}$ with $N$ nodes is defined by the Hamiltonian
\begin{equation}
	\label{eq:sphmod} \mathcal{H} =\frac{1}{2} \sum_{ij} \phi_i \mathcal{L}_{ij} \phi_j
\end{equation}
subject to the global constraint that $\sum_i \phi_i^2 = N$. Here, the graph Laplacian is given by
\begin{equation}
	\label{eq:graphlap} \mathcal{L}_{ij} = -\mathcal{A}_{ij} +z \delta_{ij}
\end{equation}
$\mathcal{A}$ is the adjacency matrix of $\mathcal{G}$, and the overall shift of $z$ ensures that $\mathcal{H}$ is positive semidefinite. Alternatively, we can view this choice of the Hamiltonian as the one which gives a system of coupled oscillators with nonnegative frequencies.

\subsection{Phase transition} 
\label{sub:phase_transition}

The partition function of the model (at inverse temperature $\beta = 1/T$) is given by the expression
\begin{equation}
	\label{eq:partfunc} Z(\beta) = \int \prod_{i} d\phi_i \, e^{-\beta \mathcal{H}[\phi]} \delta \left(\sum\limits_i \phi_i^2 - N\right)
\end{equation}
Representing the delta function by a Lagrange multiplier $\lambda$ and performing the Gaussian integral over the $\phi_i$, we obtain the effective action
\begin{equation}
	\label{eq:largeN} Z(\beta) =\int d\lambda e^{-N \left[\frac{1}{2N}\text{Tr} \log \left\{\beta\mathcal{L} -2i\lambda \mathbf{1}\right\} +i\lambda \right]}
\end{equation}

As $N\rightarrow \infty$, this integral may be performed by steepest descent. The effective propagator for the $\phi$ field is given by (defining $-\mu/T = 2i\lambda$):
\begin{equation}
	\label{eq:phi_sp_prop}
	\langle\phi_i \phi_j \rangle = T (\mathcal{L} +\mu\mathbf{1})^{-1}_{ij}
\end{equation}
and the constraint that $\sum\limits_i \langle \phi_i^2\rangle = N$ gives the self-consistency condition
\begin{equation}
	\label{eq:gap_eq} N = \text{Tr} [T (\mathcal{L} +\mu\mathbf{1})^{-1} ]
\end{equation}

We may rewrite this in terms of the eigenvalues $\epsilon_\alpha$ of the graph Laplacian
\begin{equation}
	\label{eq:eigfunc_gap_eq} \frac{1}{T} ={1\over N} \sum_{\alpha} \frac{1}{\epsilon_\alpha +\mu(T)}
\end{equation}
where we have explicitly written $\mu(T)$ to emphasize that self-consistency forces $\mu$ to depend on temperature. Throughout the paper, we use indices $\alpha, \beta$ to label the modes of the Laplacian $\mathcal{L}$ with energy $\epsilon_\alpha$ and eigenvector $u^\alpha_i$ and $i,j$ to refer to sites of the lattice.

The spectrum of $z$-regular random graph Laplacians and the related problem of hopping on the Bethe lattice have been extensively studied \cite{McKay:1981dn, Chen:1974fv}. The spectrum consists of a continuum of states in the interval $[z-2\sqrt{z-1},z+2\sqrt{z-1}]$, with the density of states
\begin{equation}
	\label{eq:DOS} \rho_c(\epsilon) = \frac{z}{2\pi} \frac{\sqrt{4(z-1) - (\epsilon-z)^2}}{z^2-(\epsilon-z)^2}
\end{equation}
However, the above density of states does not include the uniform Perron-Frobenius eigenvector, which has amplitude $1/\sqrt{N}$ on each site and energy $0$. Thus, the \emph{full} density of states is
\begin{equation}
	\label{eq:fullDOS} \rho(\epsilon) = \frac{1}{N} \delta(\epsilon) + \rho_c(\epsilon)
\end{equation}
The existence of this spectral gap, and the uniqueness of the low-lying state is guaranteed by the fact that the $z$-regular random graphs have a positive Cheeger constant.

We now return to the self-consistency equation Eq. \eqref{eq:eigfunc_gap_eq} and rewrite it in terms of the density of states Eq. \eqref{eq:fullDOS}:
\begin{equation}\label{eq:selfconsistency_dos}
	\frac{1}{T} = \frac{1}{N\mu} +\int\limits_{z-2\sqrt{(z-1)}}^{z+2\sqrt{z-1}} d\epsilon\, \frac{\rho_c(\epsilon)}{{\epsilon} +\mu}
\end{equation}
The usual argument for Bose-Einstein condensation now follows: as $T\rightarrow0$, we must decrease $\mu$ in order to satisfy this sum rule. At high temperatures, the thermodynamic limit can be taken straightforwardly without paying espercial attention to the uniform state, and one can satisfy the self-consistency equation. However, since $\mu$ cannot decrease below the lowest eigenvalue of the Laplacian - as this would render the steepest descent calculation unstable - the smallest that $\mu$ can be is zero. This occurs at a critical temperature $T_c$ given by
\begin{equation}
	\label{eq:tc_def}
	{T_c} =\left( \int\limits_{z-2\sqrt{(z-1)}}^{z+ 2\sqrt{z-1}} d\epsilon\, \frac{\rho_c(\epsilon)}{\epsilon}\right)^{-1} = \frac{z(z-2)}{z-1}
\end{equation}

This result for the transition temperature was previously derived in Ref.~\onlinecite{Cassi:1990qq}. Clearly, $T_c=0$ for $z=2$, and there is no finite-temperature transition on the chain, as expected. However, $T_c$ is finite for $z>2$, and below this temperature, we cannot satisfy the self-consistent equation in the naive thermodynamic limit discussed above. Rather, we must keep track of the uniform mode in \eqref{eq:selfconsistency_dos} and take $\mu = \frac{T}{N}\left(1-\frac{T}{T_c}\right)^{-1} + O(N^{-2})$ for $T<T_c$.

For comparison, in the $O(N_f)$ model, the Lagrange multiplier $\lambda$ would become a field $\lambda_i$, which at finite $N$ on the random graph need not  be homogeneous at the large $N_f$ saddle point. However, the infinite Bethe lattice is homogeneous and so in the thermodynamic limit we expect to recover the same saddle point as in the simpler spherical model.


\subsection{Ordered phase} 
\label{sub:ordered_phase}

How does the system behave in the low temperature phase? Consider the correlation function computed in the eigenbasis of $\mathcal{L}$, namely
\begin{equation}
	\label{eq:cl_corr_mode}
	\langle \phi_\alpha \phi_\beta \rangle = \frac{T\delta_{\alpha\beta}}{{\epsilon_\alpha} +\mu}
\end{equation}
Below the critical point, the amplitude of the lowest eigenmode is given by
\begin{equation}
	\langle \phi_0 \phi_0 \rangle ={T\over\mu} = N\left(1-\frac{T}{T_c}\right)
\end{equation}
The thermal occupation of the uniform state is macroscopic, indicative of long range order in the system.
In light of this, we revisit the steepest descent calculation, but this time treat the lowest mode separately. First, integrate out the $N-1$ higher modes
\begin{eqnarray}
	\label{eq:part_func_rewrite_ordered}
	Z(\beta) &=& \int d\lambda d\phi_0\, e^{i\lambda\left(\phi_0^2- N\right)} \nonumber\\
	& & \times \int\prod_{\alpha\neq 0}d\phi_\alpha \,e^{-\frac{1}{2}\sum\limits_{\alpha\neq 0} \left( \beta \epsilon_\alpha -2i\lambda\right) \phi_\alpha^2 } \nonumber \\
	&=& \int d\phi_0 d\lambda\, e^{-N\left[ i\lambda\left(1-\frac{\phi_0^2}{N}\right) + \frac{1}{2N}\sum\limits_{\alpha\ne0} \log \left\{\beta \epsilon_\alpha -2i\lambda \right\}\right]} \nonumber
\end{eqnarray}
and now perform the $\lambda$ integral by steepest descent, giving the self-consistency equation (again $\mu/T = -2i\lambda $)
\begin{equation}
	1 = \frac{\phi_0^2}{N} + T\int_{z-2\sqrt{z-1}}^{z+2\sqrt{z-1}} d\epsilon\, \frac{\rho_c(\epsilon)}{\epsilon +\mu}
\end{equation}
For $T$ near $T_c$, $\mu \ll z-2\sqrt{z-1}$ so we expand in powers of $\mu$ and solve:
\begin{equation}
	\mu \approx \frac{1}{bT}\left( \frac{\phi_0^2}{N} + \frac{T}{T_c} - 1\right)
\end{equation}
where $b = \int_{z-2\sqrt{z-1}}^{z+2\sqrt{z-1}} d\epsilon\, \frac{\rho_c(\epsilon)}{\epsilon^2}$.

Using this solution and approximating the steepest descent value of the integrand to quadratic order in $\mu(\phi_0)$, we find that
\begin{equation}
	Z(\beta) \approx \int d\phi_0 e^{-NV_{eff}(\phi_0)}
\end{equation}
where the effective potential for $\phi_0$ is (dropping constants):
\begin{equation}
	V_{eff}(\phi_0) = \frac{1}{2bT^2} \left[\left(\frac{\phi_0}{\sqrt{N}}\right)^2 \left(\frac{T}{T_c}-1\right) +\frac12 \left(\frac{\phi_0}{\sqrt{N}}\right)^4\right]
\end{equation}

Thus, we see that the effective potential has Landau-Ginzburg form, and predicts a mean field symmetry breaking for $T<T_c$ such that $\langle\phi_0\rangle \sim \sqrt{N}$. That is, $\langle \phi_i \rangle \sim \langle \phi_0 \rangle / \sqrt{N} \sim O(1)$.


\subsection{Spatial correlations} 
\label{sub:spatial_correlations}

We now consider the spatial behavior of the Green's function $G_{ij}=\langle \phi_i \phi_j \rangle$. As this is ultimately a Gaussian theory, the only interesting correlator is the two-point function. Since we do not know the exact eigenvectors of the graph Laplacian at finite $N$, we cannot obtain $G_{ij}$ from $G_{\alpha\beta}$ directly. Instead, we work directly in the thermodynamic limit of the Bethe lattice and assume that $G_{ij}=G(|i-j|)$ is translation invariant. In the disordered phase, $G_{ij}$ satisfies the equation of motion
\begin{equation}
	\label{eq:cl_eom}
	(\mathcal{L}_{ii'}+\mu\delta_{ii'}) G_{i'j} = - \delta_{ij}
\end{equation}
Using the \emph{ansatz} $G_{ij}=Ae^{-|i-j|/\xi}$, one finds
\begin{eqnarray}
	\label{eq:cl_corr_spat}
	G_{ij}&=&\frac{2(z-1)}{(z-2)(z+\mu)+z\sqrt{(z+\mu)^2-4(z-1)}} \nonumber\\
&& \times \left( \frac{(z+\mu)-\sqrt{(z+\mu)^2-4(z-1)}}{2(z-1)} \right)^{|i-j|}
\end{eqnarray}
which, as $\mu\to 0^{+}$ at the phase transition, reduces to
\begin{equation}
	\label{eq:cl_corr_crit}
	G_{ij} = \frac{z-1}{z(z-2)}(z-1)^{-|i-j|}
\end{equation}
This agrees with the well-known correlation length $\xi = 1/\ln(z-1)$ of the Bethe lattice at criticality and corresponds to a diverging global susceptibility.

Since the uniform mode has vanishing weight in the disordered phase, it makes no contribution to the Green's function for $T>T_c$. In the low temperature phase, the uniform mode has macroscopic occupation corresponding to a non-zero value of $\langle \phi_i \rangle$. In this case, the expression \eqref{eq:cl_corr_mode}, calculated at $\mu=0$, corresponds to the \emph{connected} correlation function $G^c_{ij}=\langle \phi_i \phi_j\rangle - \langle \phi_i \rangle\langle \phi_j \rangle$ \emph{throughout} the low temperature phase. That the correlation length sticks at its critical value and the susceptibility remains divergent throughout the condensed phase is the Bethe lattice analogue of the classical Goldstone theorem, which arises quite naturally when this model is viewed from the large $N_f$ point of view.


\section{The Quantum Spherical Model}

The quantum spherical model is given by the Hamiltonian \cite{Vojta:1996gd}
\begin{equation}
	\label{eq:sphmodq} \mathcal{H} =\frac{1}{2} g \sum_i \pi_i^2 + \frac{1}{2} \sum_{ij} \phi_i \mathcal{L}_{ij} \phi_j
\end{equation}
with canonical commutation relations $[\phi_i,\pi_j] = i\delta_{ij}$, and the \emph{mean} spherical constraint:
\begin{equation}
	\sum_i \langle \phi_i^2 \rangle = N
\end{equation}

As discussed in Ref.~\onlinecite{Vojta:1996gd}, the partition function can be rewritten in functional integral form as
\begin{eqnarray}
	\label{eq:Z_q} Z(g,\beta) &=& \int d \lambda \prod_{i} d \phi_i(\tau) \, \exp\left\{-\int_0^{\beta}d\tau\, \left[ \frac{1}{2g}\sum_i\dot{\phi}_i(\tau)^2 \right.\right.\nonumber\\
	&+&\left.\left.\frac{1}{2} \sum_{ij} \phi_i(\tau) \mathcal{L}_{ij} \phi_j(\tau) +i\lambda\left(\sum\limits_i\phi_i^2(\tau) -N\right)\right]\right\}\nonumber
\end{eqnarray}
where $\tau$ is an imaginary time parameter. Transforming to frequency space and expanding in eigenvectors of $\mathcal L$,
\begin{eqnarray}
	\label{eq:Z_qfreqspace} Z(g, \beta) &=& \int d\lambda \prod_{\alpha} d\phi_\alpha(\omega) \\
	&\times& e^{-\frac12\sum\limits_{\omega, \alpha}\phi_{\alpha}(\omega) \left[\beta\left(\frac{\omega^2}{g} + \epsilon_\alpha\right) -2i\lambda\right]\phi_{\alpha}(-\omega) -i\lambda N}\nonumber
\end{eqnarray}
where we sum over discrete frequencies $\omega_n = \frac{2\pi n}{\beta}$.

The self-consistency equation is now given by
\begin{equation}
	\label{eq:gap_q} {1} = \frac{T}{N} \sum\limits_{\omega, \alpha} \frac{1}{\left(\frac{\omega^2}{g} + \epsilon_\alpha\right) +\mu}
\end{equation}
Performing the frequency summation, we obtain
\begin{equation}
	{1} = \frac{1}{N} \sum\limits_{\alpha} \frac{\sqrt{g}}{2\sqrt{\epsilon_\alpha +\mu}} \coth\left(\frac12\beta \sqrt{g(\epsilon_\alpha +\mu)}\right)
\end{equation}
This result was computed directly within the Hamiltonian formalism in Ref.~\onlinecite{Vojta:1996gd}. Note that as $g\rightarrow 0$, we recover the classical result Eq.~\eqref{eq:eigfunc_gap_eq}.

\subsection{Quantum phase transition}

At $T= 0$, the self-consistency equation reduces to
\begin{eqnarray}
	\label{eq:gap_tzero}
	\frac{1}{\sqrt{g}} &=& \frac{1}{N}\sum\limits_{\alpha} \frac{1}{2\sqrt{\epsilon_\alpha +\mu}} \nonumber\\
	&=& \frac{1}{2N\sqrt{\mu}}+\int\limits_{z-2\sqrt{z-1}}^{z+2\sqrt{z-1}} d\epsilon\, \frac{\rho_c(\epsilon)}{2\sqrt{\epsilon + \mu}}
\end{eqnarray}
Following a similar argument as in the classical case, we find that the system has a quantum critical point at the critical coupling $g_c$, defined by
\begin{equation}
	g_c^{-1/2}=\int\limits_{z-2\sqrt{(z-1)}}^{z+ 2\sqrt{z-1}} d\epsilon\, \frac{\rho_c(\epsilon)}{2\sqrt{\epsilon}}
\end{equation}
This integral is convergent, and can be done numerically to find the precise value of $g_c$. Again, for $g < g_c$ we must be careful regarding the thermodynamic limit and take
\begin{equation}
	\sqrt{\mu} = \frac{\sqrt{g}}{2N}\left(1-\sqrt{g/g_c}\right)^{-1}
\end{equation}
in order to determine the macroscopic occupation of the zero mode. As expected, the quantum spherical model undergoes a Bose condensation transition at the critical coupling $g_c$.

\subsection{Correlations} 
\label{sub:q_correlations}

The imaginary time Green's function at zero temperature is most easily computed by taking a limit from finite temperature. From the partition function Eq. \eqref{eq:Z_qfreqspace}, we find
\begin{eqnarray}
	G_{\alpha\beta}(\tau)&=&\sum_{\omega} \frac{T \delta_{\alpha\beta}}{\frac{\omega^2}{g}+\epsilon_\alpha+\mu}e^{i \omega \tau} \nonumber \\
	&\mathop{\longrightarrow}\limits_{T\to 0}& \delta_{\alpha\beta} \int \frac{d\omega}{2\pi}\, \frac{e^{i \omega \tau}}{\frac{\omega^2}{g}+\epsilon_\alpha+\mu} \nonumber \\
	&=& \delta_{\alpha\beta} \frac{\sqrt{g}}{2\sqrt{\epsilon_\alpha + \mu}} e^{-\sqrt{g(\epsilon_\alpha + \mu)}\abs{\tau}}
\end{eqnarray}
Each mode $\alpha$ decays exponentially in imaginary time at a rate corresponding to the gap to exciting that mode. Thus, the spectral response of the time-ordered propagator in real frequency space is precisely a delta function at $\omega =  \sqrt{g(\epsilon+\mu)}$.

\begin{figure}[t]
	\centering
		\includegraphics{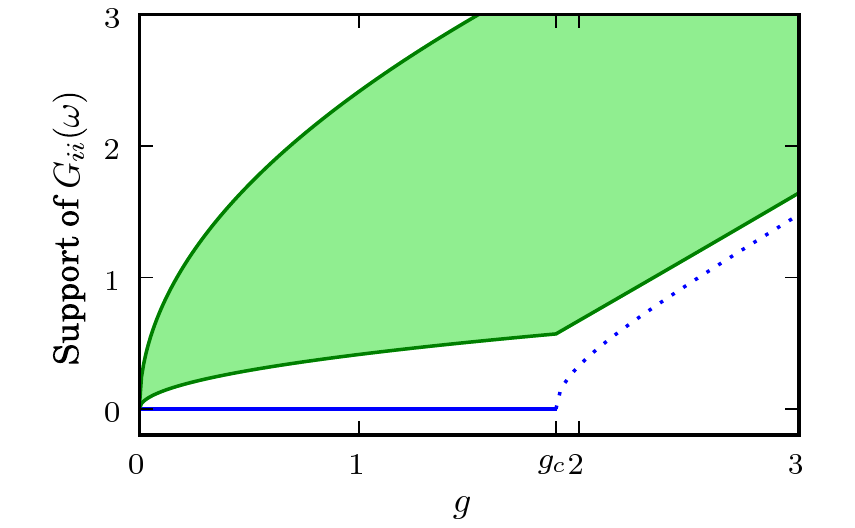}
	\caption{Support of $G_{ij}(\omega)$ at zero temperature for $z=3$. The  isolated (blue) line indicates the contribution of the uniform mode which vanishes as $1/N$ above $g_c$ (dashed) but is finite below $g_c$ signaling the long-range order (solid). Obtained from numerical inversion of Eq. \eqref{eq:gap_tzero}.}
	\label{fig:figs_fig_onsite_spectrum}
\end{figure}

Now we are in a position to understand the peculiarity of the quantum
condensation transition on the Bethe lattice. The single site Green's
function is given by transforming $G_{\alpha\beta}$ back to position space using the eigenmodes $u^\alpha_i$ of the graph Laplacian:
\begin{eqnarray}
	G_{ii}(\omega) &=& \frac{1}{N}\frac{\sqrt{g}}{2\sqrt{\mu}}\delta(\omega-\sqrt{g\mu}) \\
	&& + \sum_{\alpha>0}  u^{\alpha}_{i} u^{\alpha}_{i} \frac{\sqrt{g}}{2\sqrt{\epsilon_\alpha + \mu}} \delta(\omega -\sqrt{g(\epsilon_\alpha + \mu)}) \nonumber
\end{eqnarray}
Without detailed knowledge of the eigenvectors of the Laplacian but merely its density of states, this formula already allows us to sketch the spectral response of local single particle excitations as in Fig.~\ref{fig:figs_fig_onsite_spectrum}. As expected, the quantum phase transition is signaled by the continuous closing of a spectral gap to a uniform mode. The unusual feature is that this state is isolated from the remainder of the spectrum and its weight vanishes as $1/N$ in the disordered phase. Meanwhile, in the ordered phase, the weight
\begin{equation}
	\frac{1}{N}\frac{\sqrt{g}}{2\sqrt{\mu}} = 1-\sqrt{g/g_c}
\end{equation}
is finite, reflecting the condensation into the uniform state.

\begin{figure}[t]
	\centering
		\includegraphics{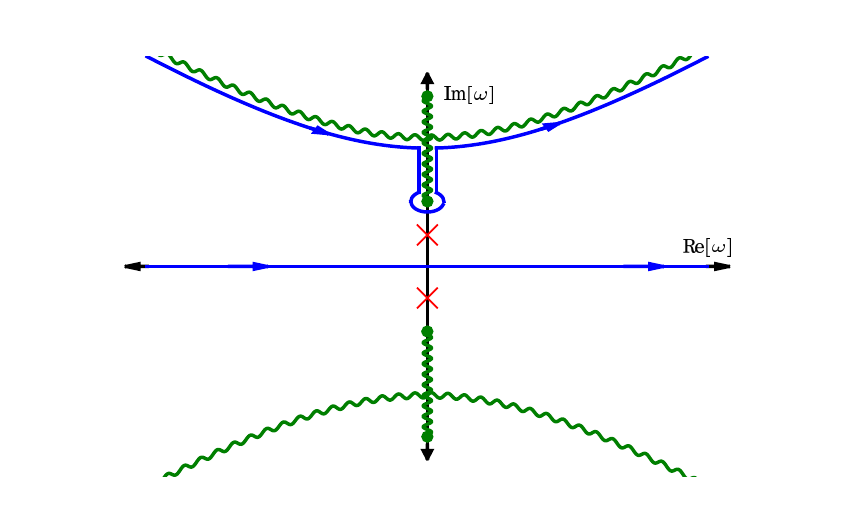}
	\caption{Analytic structure of $G_{ij}(\omega)$ in the complex plane. Wiggly lines (green) indicate branch cuts. The two crosses (red) indicate the poles due to the zero mode with residue $1/N$. The following formulae for cut locations are given at $\mu=0$ (criticality): The hyperbolic cuts are given by $\Re[\omega]^2-\Im[\omega]^2=-zg$. The vertical cuts extend between $\pm i\sqrt{g(z\pm2\sqrt{z-1})}$. These cuts control the slowest decay since we can deform the contour past the uniform mode pole in the thermodynamic limit.}
	\label{fig:polestrucGreensfunc}
\end{figure}

The general spectral features of correlations sketched above may be supplemented in detail by assuming homogeneity of the infinite Bethe lattice and exploiting our knowledge of the classical spherical model. The frequency resolution of the zero temperature Green's function $G_{ij}(\omega)$ is given by the mode sum
\begin{equation}
	\label{eq:q_corr_spat}
	G_{ij}(\omega) = \sum_{\alpha} \frac{ u_i^\alpha u_j^\alpha }{\frac{\omega^2}{g}+\epsilon_\alpha + \mu}
\end{equation}
which has the same form as the mode sum for a classical model at chemical potential $\mu+\frac{\omega^2}{g}$. Thus, we use the classical Green's function in Eq.~\eqref{eq:cl_corr_spat} and make this substitution for $\mu$. The formal expression is rather unenlightening but there is much information in the pole structure (see Fig.~\ref{fig:polestrucGreensfunc}). The most important feature is a pair of vertical square root branch cuts on the imaginary axis corresponding to the bulk spectrum. These never pinch off the real axis as $g$ is varied; rather, the closest branch point sets the dominant decay rate, which agrees exactly with the previous discussion.

At finite $N$ the branch cuts break into lines of poles and in addition one should recall the contribution of the uniform mode. This provides a pair of isolated simple poles at $\pm i \sqrt{g\mu}$ with residue $1/N$ in the disordered phase. As $g \to g_c^{+}$, these poles pinch off the real axis corresponding to the closing of the gap and the phase transition. In the condensed phase, these poles have non-vanishing residue indicating the long range order in imaginary time.

Finally, we note that the static susceptibility $\chi$ of the quantum model to a global field (applied before the thermodynamic limit) is simply given by the zero frequency lattice sum of $G_{ij}(\omega)$, which is equivalent to summing Eq. (19) over the lattice. This susceptibility diverges throughout the broken symmetry phase, as usual. However, taking a large subvolume and computing its susceptibility after the thermodynamic limit gives a convergent result, reflecting the absence of Goldstone bosons in the spectrum.

\section{Discussion}
\label{sec:Discussion}

There are two salient qualitative features of the spherical model condensation transition on the Bethe lattice that we have derived above: (1) despite its vanishing weight in the response to local excitations, there is a global mode closing the many-body gap in the disordered phase. Alternatively, this can be viewed as a `midgap' mode driving the transition, where by midgap we mean in the middle of the gap to local excitations. (2) It is via macroscopic occupation of this mode that long range order develops in the symmetry-broken phase. However, local excitations remain gapped. In fact, not only are there no Goldstone bosons, the spectral response of \emph{any} operator with bounded support will remain gapped.

We believe these features are quite general to unfrustrated transitions on the Bethe lattice and expander graphs. For one thing, as already noted in the introduction, our results are entirely consistent with those of Ref.~\onlinecite{Semerjian:2009uo} on the Bose-Hubbard model wherein the particle-hole symmetric transition is the $O(N_f)$ model at $N_f=2$. This indicates that the features we have found at $N_f = \infty$ are robust throughout the range of $O(N_f)$ models and not merely artifacts of the spherical model limit. We believe this claim will be susceptible to a proof in the $1/N_f$ expansion \cite{lpswip}. Furthermore, the results should apply to sufficiently weakly disordered ferromagnetic models, for which the disorder does not close the mode gap.

More intuitively, these results should be stable to the inclusion of self-interactions of the scalar field beyond the self-consistent Gaussian theory treated here. First, at the critical point, weak interactions cannot close the gap between the soft global modes and other massive modes; indeed, they typically cause the low energy density of states to decrease even further. Second, the usual heuristic argument for a Goldstone mode is seen to lead to a gap as follows: consider the simplest case of an $O(2)=U(1)$ broken symmetry in the broken phase.
In the usual fashion we derive an effective Lagrangian for the phase fluctuations in the symmetry-broken phase by freezing out modulus fluctuations in $|\langle  \phi  \rangle|$:
\begin{equation}
  L_{\textrm{eff}} \approx
  \frac{1}{2g}|\langle \phi \rangle|^2 \sum_i \dot{\theta_i}^2
  + \frac{1}{2} |\langle \phi \rangle|^2 \sum_{ij} \theta_i \mathcal{L}_{ij} \theta_j
\end{equation}
This is a quadratic action for a massless free field $\theta$, which would lead to gapless behavior of local excitations \emph{were the Laplacian gapless}. However, the Laplacian spectrum on expander graphs is gapped to all modulated excitations. While the global rotation is still, necessarily, of zero energy, it no longer follows that there exist local excitations of arbitrarily low energy. This corresponds to a breakdown of our intuitive understanding of generalized rigidity. Quite generally then, we conjecture that there is an `anti-Goldstone' theorem that applies to symmetry breaking transitions on expander graphs.

We should also note that the fully connected models for which mean field theory is exact exhibit similarly gapped behavior, even after the usual rescaling for extensivity. Consider the fully connected Heisenberg ferromagnet whose Hamiltonian is $H=-J/N \sum_{ij}S_i\cdot S_j = -J/N S_{tot}^2$. Trivially, the gap to the first excited state is $J$ and does not vanish in the thermodynamic limit. The fully connected graph does not have bounded coordination so is not technically an expander graph, however the surface to volume ratio is, nonetheless, very large. 

Clearly there are many subtle issues that may invalidate the usual physics lore when models are considered on non-Euclidean graphs.  As quantum models are studied more extensively in the context of quantum complexity theory and on random networks, an enhanced understanding of universal symmetry breaking properties will potentially play an important role in their analysis.

\appendix

\section*{Acknowledgments}
We thank M. Aizenman for many helpful discussions and for pointing us toward the Cheeger bound.

\bibliographystyle{apsrev-nourl}
\bibliography{Spherical-papers}

\begin{thebibliography}{16}
\expandafter\ifx\csname natexlab\endcsname\relax\def\natexlab#1{#1}\fi
\expandafter\ifx\csname bibnamefont\endcsname\relax
  \def\bibnamefont#1{#1}\fi
\expandafter\ifx\csname bibfnamefont\endcsname\relax
  \def\bibfnamefont#1{#1}\fi
\expandafter\ifx\csname citenamefont\endcsname\relax
  \def\citenamefont#1{#1}\fi
\expandafter\ifx\csname url\endcsname\relax
  \def\url#1{\texttt{#1}}\fi
\expandafter\ifx\csname urlprefix\endcsname\relax\def\urlprefix{URL }\fi
\providecommand{\bibinfo}[2]{#2}
\providecommand{\eprint}[2][]{\url{#2}}

\bibitem[{\citenamefont{Bethe}(1935)}]{Bethe:1935qc}
\bibinfo{author}{\bibfnamefont{H.~A.} \bibnamefont{Bethe}},
  \bibinfo{journal}{Proc. R. Soc. London, Ser. A}
  \textbf{\bibinfo{volume}{150}}, \bibinfo{pages}{552} (\bibinfo{year}{1935}).

\bibitem[{\citenamefont{Mezard and Parisi}(2001)}]{Mezard:2001p84}
\bibinfo{author}{\bibfnamefont{M.}~\bibnamefont{Mezard}} \bibnamefont{and}
  \bibinfo{author}{\bibfnamefont{G.}~\bibnamefont{Parisi}},
  \bibinfo{journal}{Euro. Phys. J. B} \textbf{\bibinfo{volume}{20}},
  \bibinfo{pages}{217} (\bibinfo{year}{2001}).

\bibitem[{\citenamefont{Georges et~al.}(1996)\citenamefont{Georges, Kotliar,
  Krauth, and Rozenberg}}]{Georges:1996hl}
\bibinfo{author}{\bibfnamefont{A.}~\bibnamefont{Georges}},
  \bibinfo{author}{\bibfnamefont{G.}~\bibnamefont{Kotliar}},
  \bibinfo{author}{\bibfnamefont{W.}~\bibnamefont{Krauth}}, \bibnamefont{and}
  \bibinfo{author}{\bibfnamefont{M.~J.} \bibnamefont{Rozenberg}},
  \bibinfo{journal}{Rev. Mod. Phys.} \textbf{\bibinfo{volume}{68}},
  \bibinfo{pages}{13} (\bibinfo{year}{1996}).

\bibitem[{\citenamefont{Laumann et~al.}(2008)\citenamefont{Laumann,
  Scardicchio, and Sondhi}}]{Laumann:2008rw}
\bibinfo{author}{\bibfnamefont{C.~R.} \bibnamefont{Laumann}},
  \bibinfo{author}{\bibfnamefont{A.}~\bibnamefont{Scardicchio}},
  \bibnamefont{and} \bibinfo{author}{\bibfnamefont{S.~L.}
  \bibnamefont{Sondhi}}, \bibinfo{journal}{Phys. Rev. B}
  \textbf{\bibinfo{volume}{78}}, \bibinfo{pages}{134424}
  (\bibinfo{year}{2008}).

\bibitem[{\citenamefont{Semerjian et~al.}(2009)\citenamefont{Semerjian, Tarzia,
  and Zamponi}}]{Semerjian:2009uo}
\bibinfo{author}{\bibfnamefont{G.}~\bibnamefont{Semerjian}},
  \bibinfo{author}{\bibfnamefont{M.}~\bibnamefont{Tarzia}}, \bibnamefont{and}
  \bibinfo{author}{\bibfnamefont{F.}~\bibnamefont{Zamponi}},
  \bibinfo{journal}{Phys. Rev. B} \textbf{\bibinfo{volume}{80}},
  \bibinfo{pages}{014524} (\bibinfo{year}{2009}).

\bibitem[{\citenamefont{Krzakala et~al.}(2008)\citenamefont{Krzakala, Rosso,
  Semerjian, and Zamponi}}]{Krzakala:2008rc}
\bibinfo{author}{\bibfnamefont{F.}~\bibnamefont{Krzakala}},
  \bibinfo{author}{\bibfnamefont{A.}~\bibnamefont{Rosso}},
  \bibinfo{author}{\bibfnamefont{G.}~\bibnamefont{Semerjian}},
  \bibnamefont{and} \bibinfo{author}{\bibfnamefont{F.}~\bibnamefont{Zamponi}},
  \bibinfo{journal}{Phys. Rev. B} \textbf{\bibinfo{volume}{78}},
  \bibinfo{pages}{134428} (\bibinfo{year}{2008}).

\bibitem[{\citenamefont{Nagaj et~al.}(2008)\citenamefont{Nagaj, Farhi,
  Goldstone, Shor, and Sylvester}}]{Nagaj:2008xe}
\bibinfo{author}{\bibfnamefont{D.}~\bibnamefont{Nagaj}},
  \bibinfo{author}{\bibfnamefont{E.}~\bibnamefont{Farhi}},
  \bibinfo{author}{\bibfnamefont{J.}~\bibnamefont{Goldstone}},
  \bibinfo{author}{\bibfnamefont{P.}~\bibnamefont{Shor}}, \bibnamefont{and}
  \bibinfo{author}{\bibfnamefont{I.}~\bibnamefont{Sylvester}},
  \bibinfo{journal}{Phys. Rev. B} \textbf{\bibinfo{volume}{77}},
  \bibinfo{pages}{214431} (\bibinfo{year}{2008}).

\bibitem[{\citenamefont{Kope\'{c} and Usadel}(2006)}]{Kopec:2006qr}
\bibinfo{author}{\bibfnamefont{T.~K.} \bibnamefont{Kope\'{c}}}
  \bibnamefont{and} \bibinfo{author}{\bibfnamefont{K.~D.}
  \bibnamefont{Usadel}}, \bibinfo{journal}{Phys. Status Solidi B}
  \textbf{\bibinfo{volume}{243}}, \bibinfo{pages}{502} (\bibinfo{year}{2006}).

\bibitem[{\citenamefont{Eggarter}(1974)}]{Eggarter:1974kx}
\bibinfo{author}{\bibfnamefont{T.}~\bibnamefont{Eggarter}},
  \bibinfo{journal}{Phys. Rev. B} \textbf{\bibinfo{volume}{9}},
  \bibinfo{pages}{2989} (\bibinfo{year}{1974}).

\bibitem[{\citenamefont{Johnston and Plechac}(1998)}]{Johnston:1998fv}
\bibinfo{author}{\bibfnamefont{D.}~\bibnamefont{Johnston}} \bibnamefont{and}
  \bibinfo{author}{\bibfnamefont{P.}~\bibnamefont{Plechac}},
  \bibinfo{journal}{J. Phys. A} \textbf{\bibinfo{volume}{31}},
  \bibinfo{pages}{475} (\bibinfo{year}{1998}).

\bibitem[{\citenamefont{Hoory et~al.}(2006)\citenamefont{Hoory, Linial, and
  Wigderson}}]{Hoory:2006qd}
\bibinfo{author}{\bibfnamefont{S.}~\bibnamefont{Hoory}},
  \bibinfo{author}{\bibfnamefont{N.}~\bibnamefont{Linial}}, \bibnamefont{and}
  \bibinfo{author}{\bibfnamefont{A.}~\bibnamefont{Wigderson}},
  \bibinfo{journal}{Bull. Am. Math. Soc.} \textbf{\bibinfo{volume}{43}},
  \bibinfo{pages}{439} (\bibinfo{year}{2006}).

\bibitem[{\citenamefont{Cassi and Pimpinelli}(1990)}]{Cassi:1990qq}
\bibinfo{author}{\bibfnamefont{D.}~\bibnamefont{Cassi}} \bibnamefont{and}
  \bibinfo{author}{\bibfnamefont{A.}~\bibnamefont{Pimpinelli}},
  \bibinfo{journal}{Int. J. Mod. Phys. B} \textbf{\bibinfo{volume}{4}},
  \bibinfo{pages}{1913 } (\bibinfo{year}{1990}).

\bibitem[{\citenamefont{Vojta}(1996)}]{Vojta:1996gd}
\bibinfo{author}{\bibfnamefont{T.}~\bibnamefont{Vojta}},
  \bibinfo{journal}{Phys. Rev. B} \textbf{\bibinfo{volume}{53}},
  \bibinfo{pages}{710} (\bibinfo{year}{1996}).

\bibitem[{\citenamefont{McKay}(1981)}]{McKay:1981dn}
\bibinfo{author}{\bibfnamefont{B.~D.} \bibnamefont{McKay}},
  \bibinfo{journal}{Linear Algebr. Appl.} \textbf{\bibinfo{volume}{40}},
  \bibinfo{pages}{203 } (\bibinfo{year}{1981}).

\bibitem[{\citenamefont{Chen et~al.}(1974)\citenamefont{Chen, Onsager, Bonner,
  and Nagle}}]{Chen:1974fv}
\bibinfo{author}{\bibfnamefont{M.-S.} \bibnamefont{Chen}},
  \bibinfo{author}{\bibfnamefont{L.}~\bibnamefont{Onsager}},
  \bibinfo{author}{\bibfnamefont{J.}~\bibnamefont{Bonner}}, \bibnamefont{and}
  \bibinfo{author}{\bibfnamefont{J.}~\bibnamefont{Nagle}}, \bibinfo{journal}{J.
  Chem. Phys.} \textbf{\bibinfo{volume}{60}}, \bibinfo{pages}{405}
  (\bibinfo{year}{1974}).

\bibitem[{\citenamefont{Laumann et~al.}()\citenamefont{Laumann, Parameswaran,
  and Sondhi}}]{lpswip}
\bibinfo{author}{\bibfnamefont{C.~R.} \bibnamefont{Laumann}},
  \bibinfo{author}{\bibfnamefont{S.~A.} \bibnamefont{Parameswaran}},
  \bibnamefont{and} \bibinfo{author}{\bibfnamefont{S.~L.}
  \bibnamefont{Sondhi}}, \bibinfo{note}{work in progress.}

\end{thebibliography}

\end{document}